\documentclass[10pt]{IEEEtran}
\usepackage{amsmath,url}
\usepackage{epsfig,amssymb,amsbsy,verbatim,array}
\usepackage{pstricks,psfrag,theorem,cite}

\let\intern=\iftrue

\def\figref#1{Fig.\,\ref{#1}}%
\def\E{\mathbb{E}}

\def\R{\mathbb{R}}

\def\Psi{\mathcal{L}}
\def\ie{{\em i.e.}}

\def\EIR{{\rm{EIR}}}
\def\dd{\mathrm{d}}

\def\one{\mathbf{1}}
\def\calK{\mathcal{K}}
\def\lp{\lambda_{\rm p}}
\def\unda{\underline{a}}
\def\upa{\overline{a}}
\def\undb{\underline{b}}
\def\upb{\overline{b}}

\newtheorem{theorem}{Theorem}
\newtheorem{definition}{Definition}



\begin{document}
\title{Mean Interference in Hard-Core Wireless Networks}
\author{Martin Haenggi, \IEEEmembership{Senior Member, IEEE} 
\thanks{Manuscript date \today. This work was partially supported by the NSF (grants CCF 728763
and CNS 1016742) and
the DARPA IT-MANET program through grant W911NF-07-1-0028.}}
\maketitle
\begin{abstract}
Mat\'ern hard core processes of types I and II are the
point processes of choice to model concurrent transmitters in CSMA networks.
We determine the mean interference observed at a node of the process
and compare it with the mean interference in a Poisson point process of the
same density. It turns out that despite the similarity of the two models, they
behave rather differently. For type I, the excess interference (relative to the
Poisson case) increases exponentially in the hard-core distance, while for
type II, the gap never exceeds 1 dB.
\end{abstract}
\section{Introduction}
\subsection{Motivation}
Most analyses of performance of large
ad hoc-type wireless networks is based on the stationary Poisson point process
(PPP) \cite{net:Haenggi09jsac}. 
However, the PPP is only an accurate model if the nodes are
Poisson distributed {\em and} ALOHA is used as the MAC scheme.
From a practical perspective, CSMA is much more important than ALOHA,
but it is significantly more difficult to analyze since concurrent transmitters
are spaced some minimum distance $\delta$ apart, which implies that
the numbers of nodes in disjoint areas are no longer independent.
The point processes used to model the transmitter set in CSMA
are the Mat\'ern hard-core processes of type I and type II. Both are based on
a parent PPP of intensity $\lp$. In the type I process, all nodes with a neighbor
within the {\em hard-core distance} $\delta$ are silenced,
whereas in the type II process, each node
has a random associated mark, and a node is silenced only if there is another node
within distance $\delta$ with a smaller mark. 
Such hard-core processes are difficult to analyze, since their probability generating
functionals do not exist (in contrast to clustered models, which are more tractable
\cite{net:Ganti09tit}).
 It has been argued in \cite{net:Hasan07twc,net:Nguyen07}
 that the nodes further away than $\delta$ can still be modeled as a PPP, which
 would make the analysis of CSMA networks fairly tractable.
 
 Our goal here is to verify the accuracy of the Poisson approximation by evaluating
 the mean interference measured at a typical node of the hard-core process.
 We shall see that only the type II process causes a level of interference comparable
 to the one in a PPP.

Other works on interference in CSMA networks include  \cite{net:Busson09inria},
where the mean interference is determined but at an arbitrary location on the plane
rather than at a node of the point process, and \cite{net:Busson09wasun}, which uses simulations 
to find empirical distributions.

\subsection{Preliminaries}
We first derive a general expression for the mean interference in networks
whose nodes are distributed as a stationary point process
$\Phi=\{x_1,x_2,\ldots\}\subset \R^2$ of intensity
$\lambda$. For the path loss function $g(x)$, it is assumed that
$\int_{\R^2} g(x)\dd x<\infty$,
Otherwise the interference is infinite a.s.~for any stationary $\Phi$.
The interference at the origin is defined as
\[ I\triangleq \sum_{x\in\Phi} h_x g(x) \,,\]
where $h_x$ is the power fading coefficient associated with node $x$. It is assumed that
$\E(h_x)=1$ for all $x\in\Phi$. Rather than measuring interference at an arbitrary location in $\R^2$,
we focus on the interference at the location of a node $x\in\Phi$, where it actually matters. Without loss of generality,
due to the stationarity of the point process, we may take the node to be at the origin $o$. So
the quantity of interest is $\E_o^!(I)$, which is
the mean interference measured at $o$, given that $o\in\Phi$, but not counting this node's signal power as
interference\footnote{$\E_o^!$ denotes the expectation with respect to the reduced Palm distribution.}.
Using the reduced second moment measure $\calK$ of the point process, we have
\cite{net:Haenggi08book}
\begin{equation}
\E_o^!(I)= \int_{\R^2} g(x)\calK(\dd x)\,.
\end{equation}

For a radially symmetric path loss function, with a slight abuse of notation denoted as $g(\|x\|)\equiv g(x)$, and an isotropic point process, a polar representation is more convenient:
\begin{align}
\E_o^!(I)&=2\pi \int_0^\infty g(r) \calK(r\dd r) \nonumber \\
 &=\lambda \int_0^\infty g(r) K'(r)\dd r \,.
 \label{interf_general}
\end{align}
The $K$-function is defined as
 $K(r)\triangleq\frac1\lambda\calK(b_o(r))$, 
where $b_o(r)$ is the ball of radius $r$ centered at the origin $o$, so
$K'(r)\dd r=\frac{2\pi}{\lambda}\calK(r\dd r)$.
A central quantity in our study is the {\em excess interference ratio} (EIR), defined as follows:
\begin{definition}
The excess interference ratio (EIR) is the mean interference measured at the typical point of a 
stationary hard-core point process of intensity $\lambda$ with minimum distance $\delta$
relative to the mean interference in a Poisson process of intensity $\lambda(r)=\lambda\one_{[\delta,\infty)}(r)$.
\begin{equation}
 \EIR\triangleq \E_o^!(I)/\E_o^!(I_{\rm PPP})\,.
 \end{equation}
 \end{definition}

\section{Mean Interference in Hard-Core Processes}
Hard-core processes have a guaranteed minimum distance $\delta$ between all pairs of points, which implies that $K(r)=0$ for $r<\delta$. 
In this section, we give tight bounds on the
mean interference for
Mat\'ern processes of type I and II.
We shall see that the Poisson approximation $K'(r)=2\pi r\one_{[\delta,\infty)}(r)$ provides a rather tight
lower bound for type II processes, while it gets increasingly loose as $\delta$ increases for type I processes.

\subsection{Mat\' ern process of type I}
\paragraph{Definition and $K$-function}
In this point process, points from a stationary parent PPP of intensity $\lp$ are retained only if they are at distance at least $\delta$ from all other points \cite{net:Matern86}. The intensity of the resulting
process is $\lambda=\lp\exp(-\lp\pi\delta^2)$, and
the $K$-function is 
\begin{equation}
  K(r)=2\pi\exp(2 \lp \pi \delta^2) \int_0^r u k(u)\dd u \,,
\label{cressieI}
\end{equation}
where
\begin{equation}
 k(u)=\begin{cases}
     0 & u<\delta  \\
     \exp(-\lp V_\delta(u)) & u\geq \delta \,
     \end{cases}
     \label{uku} 
 \end{equation}
 is the probability that two points at distance $u$ are both retained. It is easily
 verified that $K(r)\sim\pi r^2$ as $r\to\infty$, as is the case for all stationary point
 processes.
$V_\delta(u)$ is the area of the union of two disks of radius $\delta$ whose centers are separated by $u$,
given by
\[ V_\delta(u)=2\pi\delta^2-2\delta^2\arccos\left(\frac{u}{2\delta}\right)+ u\sqrt{\delta^2-\frac{u^2}4} \,,\;\;  0\!\leq\! u\!\leq\! 2\delta\,. \]
For $u>2\delta$, the union area is simply the area of the two disks, $2\pi \delta^2$.
First we derive a lower bound on $K(2\delta)$, the mean number of nodes within distance $2\delta$ of the origin
(not counting the node at the origin), normalized by the intensity.
We have from \eqref{cressieI}
\[ K(2\delta)=8\pi\delta^2\int_{1/2}^1 r \exp\Big(2\lambda\delta^2(\underbrace{\arccos r-r\sqrt{1-r^2}}_{f(r)})\Big)\dd r \,. \]
Let $c=\pi/3+\sqrt{3}/4$.
Since $f(r)\geq c-\sqrt{3} r$ for $1/2\leq r\leq 1$,
replacing the upper integration bound by $c/\sqrt{3}<1$ (where the bound on $f(r)$ becomes zero), and replacing $r$ in the integrand by $1/2$ yields the lower bound
\begin{align}
K(2\delta) &>8\pi\delta^2\exp(2c\lp\delta^2)\int_{1/2}^{c/\sqrt{3}} \frac12 \exp(-2\sqrt{3}\lp\delta^2r)\dd r\nonumber\\
&=\frac{2\pi}{\sqrt{3}\lp}\left[\exp\bigg(\lp\delta^2\Big(\frac{2\pi}{3}-\frac{\sqrt{3}}{2}\Big)\bigg)-1\right] \,.
\end{align}
Hence the number of points within distance $2\delta$ of the typical point,
normalized by the intensity, grows exponentially in $\delta^2$ and almost exponentially
in $\lp$. For the PPP, $K(2\delta)\propto \delta^2$.

Similarly, for the derivative, we have from \eqref{cressieI}
\[ K'(r)=2\pi\left(\frac{\lp}{\lambda}\right)^2rk(r)=2\pi\exp(2\lp\pi \delta^2) rk(r) \]
with $k(r)$ defined as in \eqref{uku}. In particular,
\[ K'(\delta)=2\pi\delta\exp\bigg(\lambda\delta^2\Big(\frac{2\pi}{3}-\frac{\sqrt{3}}{2}\Big)\bigg) \,,\]
which shows that the node density in the annulus of inner radius $\delta$ and outer radius $\delta+\dd r$ is higher than in the Poisson case by the factor $\exp(\lambda\delta^2(4\pi-3\sqrt{3})/6)\approx \exp(1.23\, \lambda\delta^2)$. This suggests that the interference will be significantly larger also.

\paragraph{Interference bounds}
Combining \eqref{interf_general} and \eqref{cressieI}, the mean interference is
\[ \E_o^!(I)=2\pi\lp \exp(\pi\lp\delta^2)\int_\delta^\infty g(r)r\exp(-\lp V_\delta(r))\dd r \,. \]
We split the interference into two terms, comprising the interference from the nodes
closer than $2\delta$ and further than $2\delta$, respectively: $I=I_{<2\delta}+I_{>2\delta}$.
We focus on $I_{<2\delta}$, \ie, the range $\delta\leq r\leq 2\delta$ first. In this range, $V_\delta(r)$ is increasing and concave,
thus we obtain an upper bound from a first-order Taylor expansion at $r=3\delta/2$:
Letting
\[ \unda\triangleq 2\arcsin\left(\frac34\right)-\frac{3\sqrt{7}}{8}\,;
\quad \undb\triangleq \frac{\sqrt{7}}{2} \]
we have
\begin{equation}
  V_\delta(r) < \left(\pi +\unda\right)\delta^2+\undb \delta r\,,\quad \delta<r<2\delta \,.
  \label{v_bound}
\end{equation}
Since $\unda<1/\sqrt{2}$ (but close), we could substitute $\unda$ with $\unda'=1/\sqrt{2}$ to obtain a simpler yet almost equally tight bound.
A lower bound on $V_\delta(r)$ is obtained by connecting the two points
 $V_\delta(\delta)=\delta^2(4\pi/3+\sqrt 3/2)$
 and $V_\delta(2\delta)=2\pi\delta^2$ by a straight line. 
 This yields
 \begin{equation}
 V_\delta(r) > \left(\pi +\upa\right)\delta^2+\upb \delta r\,,\quad \delta<r<2\delta \,,
  \label{v_bound_low}
\end{equation}
for
\[ \upa\triangleq\sqrt{3}-\frac{\pi}{3}\,;\quad \upb\triangleq\frac{2\pi}3-\frac{\sqrt{3}}{2}\,.\]
To use these affine bounds on $V_\delta(r)$ to bound the mean interference, we define
\begin{align}
 h(a,b) 
   &\triangleq 2\pi\lp e^{-\lp a \delta^2}\int_\delta^{2\delta} g(r)r e^{-\lp b\delta r}\dd r \nonumber \\
   &=2\pi\lp e^{-\lp a \delta^2} H(\lp b\delta,\delta)\,,
 \end{align}
 where $H(v,x)\triangleq\int_x^{2x} g(r)r\exp(-v r)\dd r$. 
Upper and lower bounds on $\E_o^!(I_{<2\delta})$ can now be expressed as:
 \begin{equation}
  h(\unda,\undb) < \E_o^!(I_{<2\delta}) < h(\upa,\upb) 
  \label{h_bounds}
 \end{equation} 
 
\begin{figure}
 \centerline{\epsfig{file=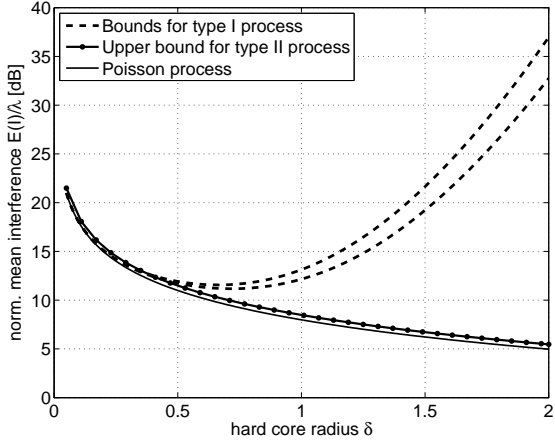,width=.85\columnwidth}}
 \caption{Normalized mean interference $\E_o^!(I)/\lambda$ for the Poisson point process
 (bottom solid curve), the upper bound from \eqref{eir_II_power} for the Mat\'ern process of type II (dotted solid curve), and
 upper and lower mean interference bound for the Mat\'ern process of type I, for $\lp=2$ and $\alpha=3$.
 The EIR (gap) between the Poisson and type II
 curves is $0.5$ dB, while the gap between the Poisson and type I curves increases exponentially with $\lp$ and $\delta$.
  At $\delta=2$, the EIR is about 30dB.}
 \label{fig:bounds}
 \end{figure}

 Specializing to the class of power path loss laws\footnote{An exponential factor in the path loss law can easily
be accommodated: The only change is in the constant $b$.} $g(r)=(\max\{r_0,r\})^{-\alpha}$, where $0\leq r_0\leq\delta$, 
 there exists a concrete expression for $H$:
 \[ H(v,x)=v^{\alpha-2}\big(\Gamma(2-\alpha,v x)-\Gamma(2-\alpha,2v x)\big)\,. \]
 \figref{fig:bounds} shows the bounds \eqref{h_bounds}, normalized by the intensity $\lambda$,
  for $\alpha=3$ and $\lp=2$, as a function of $\delta$ (dashed curves).

The interference from nodes outside $r>2\delta$ is the same as in the (equi-dense)
PPP:
\[ \E_o^!(I_{>2\delta})=2\pi\lp\frac{\exp(-\lp\pi\delta^2)}{(2\delta)^{\alpha-2}(\alpha-2)} \]
The total interference in the PPP is obtained by replacing the $2\delta$ in the denominator
by $\delta$, hence $\E_o^!(I_{\rm PPP})=2^{\alpha-2} \E_o^!(I_{>2\delta})$. 
For the excess interference ratio, we find
\begin{equation}
\EIR=\frac{1}{2^{\alpha-2}}\left(\frac{\E_o^!(I_{<2\delta})}{\E_o^!(I_{>2\delta})}+1\right) \,.
\label{eir}
\end{equation}

\begin{theorem}
For power path loss laws $g(r)$ with exponent $\alpha$,
the excess interference in the Mat\'ern process of type I grows exponentially,
{\rm \ie},
\begin{equation}
  \EIR=\Omega(e^{\lp \delta^2})\,,\quad \lp\delta\to\infty \,.
\end{equation}
\end{theorem}
\begin{IEEEproof}
Using the lower bound in \eqref{h_bounds},
\[ \EIR > e^{\lp \delta^2(\pi-\unda)} H(\lp \undb\delta,
\delta)(2\delta)^{\alpha-2}(\alpha-2) \,. \]
Since $H(v,x)\sim (vx^{\alpha-1}e^{vx})^{-1}$ as $\lp\delta\to\infty$,
\[ \EIR =\Omega\left(\frac{e^{\lp \delta^2(\pi-\unda-\undb)}}{\lp \delta^2}\right) \,,\quad\lp\delta\to\infty\,.\]
The result follows from $\pi-\unda-\undb>1$.
\end{IEEEproof}

Keeping track of the pre-constants, we obtain an approximation, quite accurate for $\lp\delta^2>4$:
\begin{equation}
  \EIR \approx \frac{(\alpha-2) 2^{\alpha-2} e^{\lp \delta^2(\pi-\unda-\undb)}}{\lp \undb\delta^2} 
\end{equation}
For the parameters in \figref{fig:bounds}, at $\delta=2$, this yields 31.5dB.

 \subsection{Mat\'ern process of type II}
 Here, a random mark is associated with each point, and a point of the parent Poisson
 process is deleted if there exists another point within the hard-core distance $\delta$
 with a smaller mark. The intensity of the resulting process is \cite{net:Matern86}
 \[ \lambda=\frac{1-\exp(-\lp\pi\delta^2)}{\pi\delta^2} \]
 and the probability that two points at distance $r$ are retained is, for $r\geq \delta$,
 \[ k(r)=\frac{2V_\delta(r)(1-e^{-\lp\pi\delta^2})-2\pi\delta^2(1-e^{-\lp V_\delta(r)})}
 {\lp^2\pi\delta^2 V_\delta(r)(V_\delta(r)-\pi\delta^2)} \,. \]

 \begin{theorem}
Irrespective of the path loss function $g(r)$ and all other parameters,
the excess interference ratio for Mat\'ern processes of type II never exceeds 
 \begin{equation}
  \nu\triangleq\frac{12\pi}{8\pi+3\sqrt 3} < \frac54 < 1 {\rm dB}\,.
 \end{equation} 
 For power path loss laws with exponent $\alpha$, the bound can be sharpened to
 \begin{equation}
  \nu-\frac{\nu-1}{2^{\alpha-2}} \,.
  \label{eir_II_power}
  \end{equation}
 \end{theorem}
 \begin{IEEEproof}
 First we note that $(\frac{\lp}{\lambda})^2 k(r)$ is monotonically increasing
 in $\lp$ and $\delta$ for all $\delta\leq r<2\delta$. For $r\geq 2\delta$, we have
$(\frac{\lp}{\lambda})^2 k(r) \equiv 1$, since outside distance $2\delta$ the hard-core
process behaves like a PPP. This implies that the EIR can only increase with 
$\lp$ and $\delta$ (which is intuitive, since for $\lp\to 0$ or $\delta\to 0$, the process
is Poisson). Hence letting $\lp\delta\to\infty$ yields an upper bound on the EIR. We have
\[ k(\delta)\sim  \frac{2}{\lp^2\pi\delta^4c} \,,\quad c\triangleq 4\pi/3+\sqrt 3/2\,, \]
which upper bounds $k(r)$ for all $r\geq\delta$ and all finite $\lp$ and $\delta$.
Consequently,
 \[ \frac{\E_o^!(I_{<2\delta})}{\lambda} < \int_\delta^{2\delta} 2\pi g(r)\frac{\lp^2}{\lambda^2} \frac{2}{\lp^2\pi\delta^4c} r\dd r=
\int_\delta^{2\delta} g(r)\frac{4\pi^2}{c}r \dd r\,, \]
where the RHS is $\nu=2\pi/c'$ times the mean interference in the Poisson case.
Inserting this bound into \eqref{eir} yields the result for the power path loss law.
 \end{IEEEproof}
 For $\alpha=3$, this is quite exactly 0.5 dB, as reflected in \figref{fig:bounds}.
 
 \section{Conclusion}
 The behavior of two popular point process models for CSMA networks differs greatly.
 For the Mat\'ern hard-core process of type I, the excess interference relative to the Poisson
 point process increases exponentially in the parent process density $\lp$ and the hard-core
 distance $\delta$ (for power path loss laws), while for Mat\'ern processes of type II, the excess interference never
 exceeds 1dB, irrespective of the path loss law.

\end{document}